# Localized modes in dissipative lattice media: An overview

Yingji He[1], Boris A. Malomed[2] and Dumitru Mihalache[3,4]

[1]*School of Electronics and Information, Guangdong Polytechnic Normal University, 510665 Guangzhou, China*

[2]*Department of Physical Electronics, Faculty of Engineering, Tel Aviv University, Tel Aviv 69978, Israel*

[3]*Academy of Romanian Scientists, 54 Splaiul Independentei, RO-050094, Bucharest, Romania*

[4]*Horia Hulubei National Institute for Physics and Nuclear Engineering, P.O.B. MG-6, RO-077125 Magurele-Bucharest, Romania*

E-mail addresses: (1) heyingji8@126.com; (2) malomed@post.tau.ac.il; (3) Dumitru.Mihalache@nipne.ro

## Abstract

We overview recent theoretical studies of the dynamics of one- and two-dimensional spatial dissipative solitons in models based on the complex Ginzburg-Landau equations with the cubic-quintic combination of loss and gain terms, which include imaginary, real, or complex spatially periodic potentials. The imaginary potential represents periodic modulation of the local loss and gain. It is shown that the effective gradient force, induced by the inhomogeneous loss distribution, gives rise to three generic propagation scenarios for one-dimensional (1D) dissipative solitons: transverse drift, persistent swing motion, and damped oscillations. When the lattice-average loss/gain value is zero, and the real potential has spatial parity opposite to that of the imaginary component, the respective complex potential is a realization of the parity-time symmetry. Under the action of lattice potentials of the latter type, 1D solitons feature unique motion regimes in the form of transverse drift and persistent swing. In the 2D geometry, three types of axisymmetric radial lattices are considered, *viz*., ones based solely on the refractive-index modulation, or solely on the linear-loss modulation, or on a combination of both. The rotary motion of solitons in such axisymmetric potentials can be effectively controlled by varying the strength of the initial tangential kick.

# 1. Introduction

Recently, the evolution of nonlinear localized excitations in dissipative media including lattice structures, such as the ones generated by spatially periodic modulation of the local loss/gain coefficient, has been widely paid more attention. In particular, it has been shown that dissipative solitons may be strongly affected by the spatially inhomogeneous gain [1-4]. In this context, it is relevant to mention that spatially inhomogeneous gain landscapes can support stable dissipative solitons [1,2], defect modes [3], multi-peak patterns [4], and other localized modes. One and two-dimensional (1D and 2D) dissipative bistable systems with spatially periodic forcing also exhibit localized structures, as was shown both theoretically and experimentally [5]. Stable nonlinear patterns in Bose-Einstein condensates loaded in dissipative optical lattices (OLs) with nonlinear losses were studied too [6]. Moreover, in dissipative systems based on the 2D complex Ginzburg-Landau (CGL) model with the cubic-quintic (CQ) nonlinearity, various stable vortical 2D dissipative solitons in the presence of radially inhomogeneous losses [7], and 1D dissipative solitons, which perform controlled drift and swing motion in the presence of periodically inhomogeneous losses [8], were found too. Specific complex-valued potentials composed of loss/gain and refractive-index lattices, satisfying the condition of the parity-time (*PT*) symmetry, have recently drawn a great deal of interest due to the principal possibility to realize non-Hermitian Hamiltonians in such systems, and emerging applications to various physical settings [9]. Particularly, such *PT*-symmetric potentials can support stable optical solitons with new properties, concerning their existence and stability, in 1D and 2D optical systems [10-17].

Defect modes in *PT*-symmetric periodic complex-valued potentials have been studied too [18]. Rich dynamics of spatial dissipative solitons in the CQ CGL model with *PT*-symmetric periodic potentials, including controllable drift and swing motion, were also revealed by the analysis of the corresponding models [19].

Besides of the regular lattices, cylindrical OLs, such as Bessel OLs, can support rotary motion of 2D solitons in the corresponding annular potential troughs [20,21]. Bessel OLs also provide for stabilization of three-dimensional (3D) spatiotemporal optical solitons [22]. The radial Bessel potential maintains ring solitons in the form of azimuthal dipoles and quadrupoles, if the nonlinearity is repulsive [23], and a potential periodic in the radial direction gives rise to radial gap solitons [24]. Optical spatial solitons trapped either at the center of the lattice or in the Bessel-like ring lattices were created in the experiment [25]. Recently, the rotary dynamics of dissipative spatial solitons was studied in the CQ CGL model with three kinds of periodic cylindrical lattices, which are generated by the refractive-index modulation, or linear-loss modulation, or a combination of both [26].

In this article we present an outline of basic theoretical results obtained for lattice solitons which can be supported in the CQ CGL model. It is a universal model that plays an important role in many areas such as superconductivity and superfluidity, fluid dynamics, reaction-diffusion phenomena, nonlinear optics, Bose–Einstein condensation, quantum field theory, and biology and medicine [27-31].

The most general model, including various lattices controlling the spatial-beam propagation, in terms of the optical realization, is described in Sect. 2. Then, in Sect. 3 we report properties of dissipative spatial solitons in the presence of sinusoidally-modulated spatially inhomogeneous losses. In Sect. 4, we consider dissipative spatial solitons in the presence of *PT*-symmetric optical lattices, which are described by complex-valued periodic potentials. The dynamics of dissipative spatial solitons in the presence of three kinds of periodic cylindrical lattices, which are generated by the refractive-index modulation, or linear-loss modulation, or the combination of both, are briefly reviewed in Sect. 5. The paper is concluded by Sect. 6.

## 2. The governing model

We consider the generic 2D CQ CGL equation with a periodic lattice potential [26]:

$$iu_z + (1/2)(u_{xx} + u_{yy}) + |u|^2 u + \nu |u|^4 u = iN[u] - [R(x,y) - iL(x,y)]u, \quad (1)$$

where $z$ is the propagation distance, and $(x, y)$ are the transverse coordinates. In the case of a periodic axisymmetric lattice potential representing the cylindrical lattice, one has $R(x, y) = R(r)$ and $L(x, y) = L(r)$, where $r = \sqrt{x^2 + y^2}$. Further, $\nu$ is the quintic self-defocusing coefficient, and the combination of the CQ nonlinear terms is $N[u] = -\alpha u + \varepsilon |u|^2 u + \mu |u|^4 u$, where $\alpha$ is the linear-loss coefficient, $\mu<0$ is the quintic-loss parameter, and coefficient $\varepsilon$ accounts for the cubic gain. In the case of the 1D version of the 2D CQ CGL (1) the third term in the left-hand side of Eq. (1) is dropped, and the 1D periodic potential is taken as $R(x, y) = R(x)$ and $L(x, y) = L(x)$ [8,19]. In the following, we consider a generic case, which can be adequately represented by a set of parameters with $\nu = -0.2$, $\mu = -1$, and $\varepsilon = 1.6$. Equation (1) can be implemented directly as a model of spatially patterned laser cavities [7,8].

## 3. Soliton dynamics induced by periodic spatially inhomogeneous losses in media described by the complex Ginzburg–Landau model, as per Ref. [8]

Following Ref. [8], we consider the 1D version of Eq. (1) and the loss-modulation function of the sinusoidal form, $L(x) = d \sin(x/5)$, while $R(x) = 0$. The profile of $L(x)$ is shown in Fig. 1(a), the input soliton being launched at $x = 0$. Then we vary linear loss coefficient $\alpha$ and amplitude $d$ of the inhomogeneous loss in order to study the ensuing soliton dynamics. Different domains of the corresponding soliton propagation scenarios are plotted in Fig. 1(b). We see that the linear loss coefficient, $\alpha$, increases with the growth of amplitude $d$ of the inhomogeneous loss profile. For certain amplitudes of the modulated loss, solitons spread in the course of the propagation, due to excess gain, when coefficient $\alpha$ is smaller than a critical value [below curve 1 in Fig. 1(b)], as shown in Fig. 1(c). If $\alpha$ increases up to the region between curves 1 and 2, the transverse gradient force produced by the inhomogeneous loss induces a leftward drift of the soliton, see Fig. 1(d). From Fig. 1(d) we see that the soliton may even move with acceleration at the initial stage of propagation, due to the power loss suffered when the soliton passes dissipative channels of the structure. At small values of $\alpha$, the velocity of the soliton's drift is large, causing it to quickly move from the input position. At large $\alpha$, the soliton first exhibits an expanding swing over a long propagation distance, and then drifts far away from the input position. When the linear-loss coefficient increases to values corresponding to curve 2 of Fig. 1(b), the solitons perform a persistent swing, see Fig. 1(e). Curve 2 in Fig. 1(b), which corresponds to the transition between the drift and oscillatory localized states, belongs to the saddle-node instability, as the potential energy has a maximum value there.

If $\alpha$ exceeds the values corresponding to curve 2, the solitons first drift to the right over a short spatial interval, which is limited to half a period of the periodic loss-modulation function $L(x)$, which is followed by damped oscillations [Fig. 1(f)]. Eventually, this swinging soliton transforms into a stationary one, located to the right of the input position ($x = 0$), as seen in Fig. 1(f), where the output position of the new soliton is $x_{\text{out}} \approx 2.3$. When $\alpha$ exceeds the critical value corresponding to curve 3 ($\alpha_{cr} = 0.54$), the soliton decays under the action of heavy losses [Fig. 1(g)].

The drift and swing of solitons arise as a consequence of the gradient force induced by the spatially inhomogeneous loss. In fact, the inhomogeneity of the loss breaks the gain-loss balance of the dissipative soliton, causing the soliton to drift to an appropriate position in order to maintain its stability. In the course of drifting, the profile of the dissipative soliton keeps changing due to the combined effects of the gain and loss, and of the diffraction and self-focusing nonlinearity. Thus, if the drifting soliton cannot find an appropriate position near the input channel, it will drift continually; in contrast, if the soliton can find such a position near the input channel, it will be either swinging around this position, or will gradually cease swinging at an appropriate equilibrium position. The actual shape of the drifting localized state is asymmetric, due to the effect of the inhomogeneous loss-modulation profile.

If a stochastic term is added to the model, ratchet motion of localized modes is expected, provided that the loss-modulation profile is given an asymmetric profile along axis *z*. Then, all regions of the dynamical behavior of the localized modes will be modified accordingly.

It is relevant to mention too that the boundaries of the integration window hardly affect the results, because the computation domain is large enough in all figures.

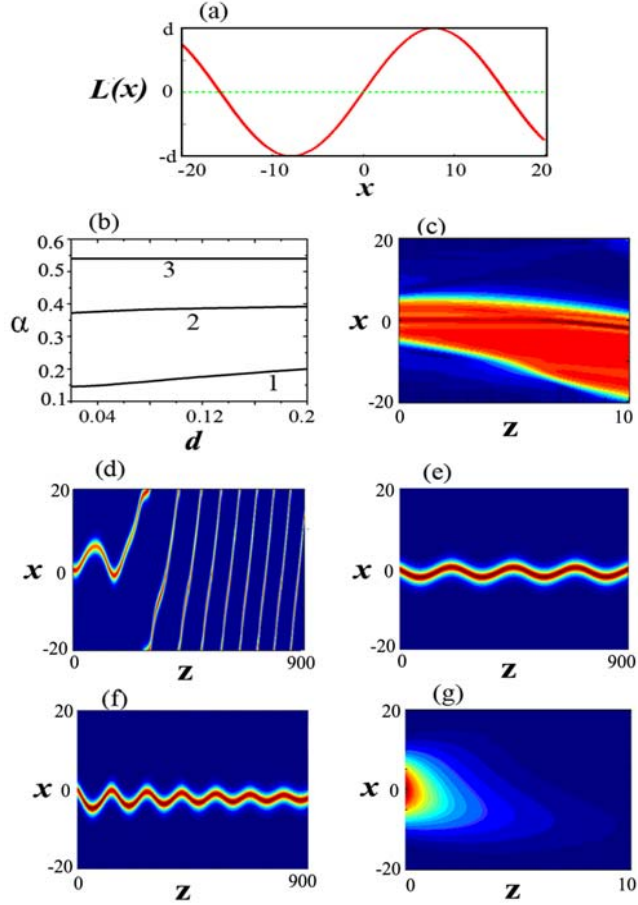

**Figure 1.** (Color online) (a) The inhomogeneous loss-modulation profile, $L(x)=d \sin(x/5)$. (b) Soliton dynamics described by the relation of the linear-loss coefficient, $\alpha$, to the amplitude $d$ of the inhomogeneous loss structure. The so represented dynamical regimes include spreading out under the action of the excess gain (below curve 1), leftward drift (between curves 1 and 2), persistent swing (on curve 2), damped oscillations (between curves 2 and 3), and decay (above curve 3). Examples of the simulated evolution of the solitons: (c) spreading out under the excess gain, at $\alpha=0.1$ and $d=0.2$; (d) the leftward drift at $\alpha=0.36$ and $d = 0.2$; (e) persistent swing at $\alpha=0.385$ and $d = 0.1$; (f) damped oscillations at $\alpha=0.45$ and $d = 0.2$; (g) decay at $\alpha=0.56$ and $d=0.2$. The results are presented as per Ref. [8].

## 4. Lattice solitons in media described by the complex Ginzburg-Landau model with *PT*-symmetric periodic potentials, as per Ref. [19]

Following Ref. [19], the last term in the 1D version of Eq. (1) may be taken in the form of a *PT*-symmetric linear OL. To present typical examples, we focus here on the periodic potentials with $R(x) = A_1 \cos^2(x/T_1)$ and $L(x) = -A_2 \sin(x/T_2)$, where $A_1$ and $A_2$ are amplitudes of real and imaginary parts of the *PT*-symmetric lattice potential, respectively, $\pi T_1$ and $2\pi T_2$ being the corresponding periods.

Figure 2(a) shows domains in the plane of $(\alpha, T_2)$, corresponding to different scenarios of the soliton dynamics, for fixed period $T_1 = 1$ of the real part of the *PT*-symmetric potential. This relatively small value of $T_1$ corresponds to a tightly bound OL. Note that the real part of the *PT*-symmetric OL represents the spatial modulation of the refractive index, while the imaginary part of the *PT*-symmetric potential represents the spatially inhomogeneous loss, which in turn can produce a transverse gradient force. The magnitude of this force is determined by the slope of the inhomogeneous loss profile. Note that, for the chosen lattice period, $T_1 = 1$, the size of a single lattice cell approximately matches the soliton's width; hence such OLs can tightly bound the spatial dissipative soliton. From Fig. 2(a), we see that the spatial soliton exhibits four distinct dynamical scenarios, corresponding to (a) spreading out under the excess gain (region A), (b) drift to an adjacent lattice position (B), (c) stable straight propagation (C), and (d) decay of the soliton [the top strip in Fig. 2(a) for $\alpha > 0.54$]. From Fig. 2(a) we see that, with the increase of period $T_2$ of the imaginary part of the *PT*-symmetric OL, region B, which corresponds to the drift of the soliton towards an adjacent lattice site, gradually shrinks until $T_2 \approx 1$, whereas for $T_2 > 1$ region B disappears because in this situation the gradient force of the inhomogeneous loss is too small to power the drift. We have found that, for relatively small values of $T_2$ in the interval $0.5 < T_2 < 1$, the gradient force induced by the inhomogeneous loss can push the soliton to the other lattice site, and stationary propagation takes place in that lattice channel. This happens because the gain-loss balance, needed for the formation of a stable soliton, broke down, and the soliton must seek a new appropriate channel for the stable propagation. As a typical example, we take $T_2 = 0.55$ and analyze the effect of the linear loss coefficient, *α*, on soliton propagation by fixing the period as $T_1 = 1$.

If *α* is small, the soliton can spread out upon the propagation due to the excess gain [region A in Fig. 2(a)]. An example of the propagation affected by the excess gain, at $\alpha = 0.2$ and $T_2 = 0.55$, is shown in Fig. 2(c). The excess gain amplifies both the soliton and background noise. When $\alpha$ increases further, the soliton can be pushed to another lattice channel and stably propagate in it [region B in Fig. 2(a)]. The transverse displacement from the original position decreases with the increase of *α*, as seen in Fig. 2(d) for $\alpha = 0.25$. On the other hand, the soliton is tightly bound at the input lattice position, and stably propagates in this way, if $\alpha$ is large enough [region C in Fig. 2(a)]. Figure 2(e) shows an example of the stationary propagation for $\alpha$, large enough, *viz.*, $\alpha = 0.5$. However, when *α* exceeds a critical value, $\alpha > 0.54$, the soliton quickly decays, as shown in Fig. 2(f). Additional numerical simulations show that, for fixed lattice period, $T_1 = 1$, the soliton-drift scenario also occurs at $T_2 \leq 0.5$.

Next, we analyze the soliton propagation dynamics for large values of the period of the real part of the *PT*-symmetric potential, $T_1 > 1$. In this case, the soliton can swing sideways or drift within several lattice periods, a situation which is very different from the generic propagation scenarios displayed in Figs. 2(c)-(e) for $T_1 = 1$. First, we select a small value of $T_2$ (here we take $T_2 = 0.5$) and we consider $T_1 > 1$. In Fig. 2(b), we show the dependence of the linear-loss coefficient, $\alpha$, on period $T_1$ of the real part of the *PT*-symmetric potential. Four domains in the plane of ($\alpha$, $T_1$) correspond to the following propagation scenarios: spreading out due to the excess gain (region A), soliton's drift (B), persistent swing (C), and decay of the soliton (for $\alpha > 0.54$). For small values of $\alpha$, the input soliton spreads over several lattices due to the presence of the excess gain. When $\alpha$ increases further, the input soliton can run across the lattices, and can drift in the transverse direction, see Fig. 2(g). If $\alpha$ is large enough, the input soliton performs a persistent swing motion around its original channel, as shown in Fig. 2(h). The swinging amplitude decreases with the increase of $\alpha$. The soliton quickly decays when $\alpha > 0.54$.

At $T_1 > 1$, the lattice channel is wide. In this case, the beam spreads at the initial stage of the propagation. Then, the spreading is arrested by the lattice potential, and the beam maintains stable propagation.

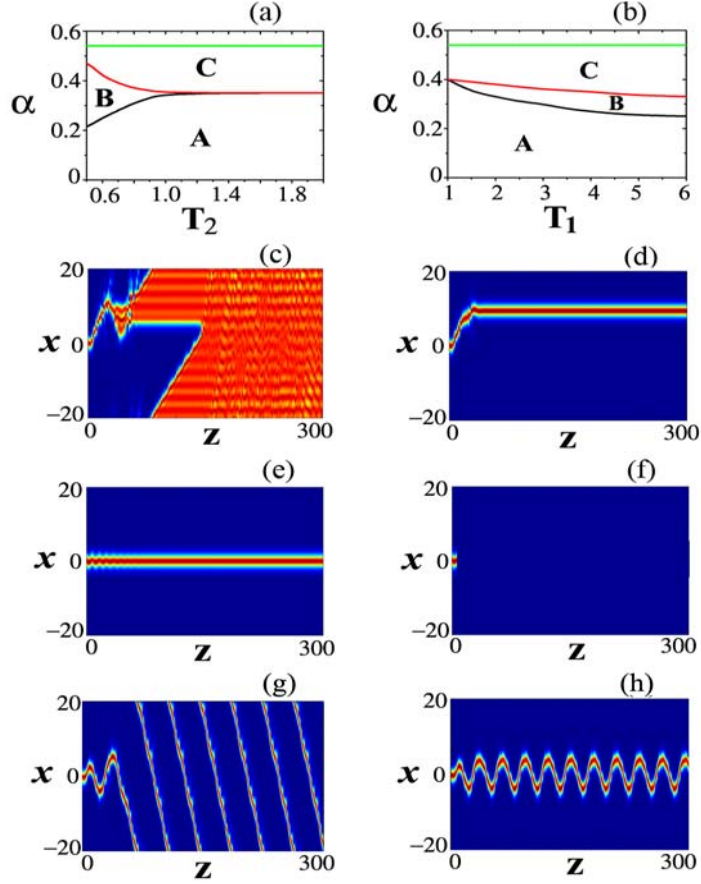

**Figure 2.** (Color online) (a) The dependence of the linear loss coefficient, $\alpha$, on period $T_2$ of the imaginary part of the *PT*-symmetric periodic potential, with a relatively small period, $T_1 = 1$, of the real part. Domains of different propagation scenarios: decay under the action of the excess gain (A), soliton's drift to an adjacent lattice site (B), stable straight propagation (C), and decay of the soliton, at $\alpha > 0.54$. (b) The dependence of $\alpha$ on $T_1$ in the region of $T_1 > 1$. Regions of different propagation scenarios: the spreading out under the action of the excess gain (A), soliton's drift (B),

persistent swing (C), and decay of the soliton, at $\alpha > 0.54$. Examples of the evolution scenarios: (c) spreading out due to the excess gain at $\alpha = 0.2$ and $T_2 = 0.55$; (d) soliton's drift at $\alpha = 0.25$ and $T_2 = 0.55$; (e) straight propagation for $\alpha = 0.5$ and $T_2 = 0.55$; (f) decay at $\alpha = 0.55$ and $T_2 = 0.55$; (g) soliton's drift at $\alpha = 0.3$, $T_1 = 4$, and $T_2 = 0.5$; (h) persistent swing at $\alpha = 0.4$, $T_1 = 4$, and $T_2 = 0.5$. Other parameters are $A_1 = 0.2$, and $A_2 = 0.2$. The results are presented as per Ref. [19].

## 5. Rotary dissipative spatial solitons in cylindrical lattices, as per Ref. [26]

Following Ref. [26], we focus here on three kinds of cylindrical lattices: **(i)** only with the refractive-index modulation, $R(r) = R_0 \cos^2(r)$ and $L(r) = 0$; **(ii)** only with the linear-loss modulation, $R(r) = 0$ and $L(r) = -L_0 \sin(2r)$; **(iii)** a combination of both types of the modulation: $R(r) = R_0 \cos^2(r)$ and $L(r) = -L_0 \sin(2r)$. Here $R_0$ and $L_0$ are amplitudes of the two types of the modulation, and the linear-loss coefficient in Eq. (1) is fixed to be $\alpha = 0.4$.

### (a) The axisymmetric lattice with the refractive-index modulation

We first consider the lattice with the refractive-index modulation of the form $R(r) = 3\cos^2 r$. In the stationary state, the soliton can stably exist in the circular trough of the lattice, see Fig. 3(a). To initiate the dynamics, we imprint a non-uniform phase on the input soliton, multiplying the input field by $\exp(iMy)$, where the kick strength, $M$, is proportional to the initial momentum imparted to the soliton in the tangential direction. The soliton kicked tangentially in this way can perform rotary motion along the circular lattice ring, provided that the kick takes values below a maximum one, $M_{max}$, see Fig. 3(c), while the rotary motion is unstable at $M > M_{max}$. In the latter case, the soliton moves across the lattice and quickly decays [Fig. 3(d)]. The value of $M_{max}$ is shown versus the refractive-index modulation depth, $R_0$, in Fig. 3(b). Note that, to maintain the power balance, the cubic gain must compensate the linear loss when soliton propagates in the medium described by the model (1), but the cubic gain hardly affects $M_{max}$.

### (b) The axisymmetric lattice with the linear-loss modulation

Next we consider the annular lattice in which solely the linear-loss is subject to the spatial modulation, in the form of $L(r) = -L_0 \sin(2r)$. Similar to the case discussed above, this modulation profile can also support the soliton's rotary motion in the circular trough, and there also exists a largest kick's strength, $M_{max}$, which bounds the unstable and stable rotary motions. When the input soliton is located in the first annular trough of the lattice, see Fig. 3(e), the dependence of $M_{max}$ on depth $L_0$ of the loss-modulation profile is shown in Fig. 3(f), which demonstrates that the stable rotary motion of the soliton takes place for values of the loss-modulation depth from interval $0.25 \leq L_0 \leq 0.8$. It is seen too that $M_{max}$ is not a monotonic function; at $L_{0,max} = 0.68$, $M_{max}$ attains its largest value, $M_{max}$ decreasing beyond this point with the increase of $L_0$.

At $M > M_{max}$, the soliton runs across the lattice as in the case of the refractive-index modulation, considered above. However, in the present case the soliton escaping from the annular lattice evolves into a disordered mode, due to the effect of the inhomogeneous loss-modulation profile, see Fig. 3(h).

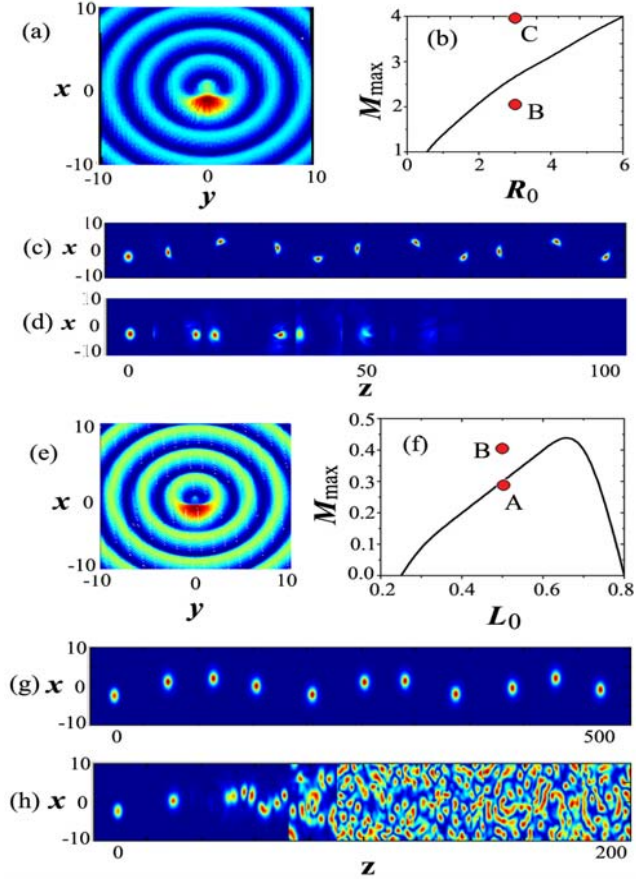

**Figure 3**. (Color online) (a) The soliton's rotary motion in the first annular trough of the cylindrical lattice with only refractive-index modulation: $R(r) = 3\cos^2 r$ and $L(r) = 0$. (b) The maximum value $M_{max}$ of the tangential kick versus the refractive-index modulation depth, $R_0$. Here and in panel (f) below, the rotary motion is unstable above the solid curve. (c) and (d) Stable rotary motion for $M = 2$, and unstable motion for $M = 4$, respectively, corresponding to points B and C in (b). (e) The soliton's rotary motion in the first annular trough of the lattice with only linear-loss modulation: $L(r) = -L_0\sin(2r)$ and $R(r) = 0$. (f) The maximum value of the initial tangential kick, $M_{max}$, versus the linear-loss modulation depth, $L_0$. **(g) and (h):** Stable rotary motion for $M = 0.3$ and $L_0 = 0.5$, and unstable motion for $M = 0.4$ and $L_0 = 0.5$, **respectively,** corresponding to points A and B in (f). The results are presented as per Ref. [26].

**(c) Radial symmetric lattices with combined modulations of the refractive index and local loss**

Finally, we study the dynamics of the dissipative spatial solitons under combined refractive-index and loss modulation profiles. First, we have found that stationary solitons, (for $M = 0$) can stably propagate in such complex cylindrical lattices. Next, we consider the possibility of the rotary motion of the input solitons in this setting. We have found that, for a fixed value of amplitude $R_0$ of the refractive-index modulation profile, the maximum tangential-kick's strength, $M_{max}$, decreases with the increase of amplitude $L_0$ of the loss-modulation profile, as shown in Fig. 4(a). Also, we have found that, for fixed amplitude $L_0$, $M_{max}$ increases with the increase of $R_0$, as shown in Fig. 4(b). However, in the case of stable evolution, $M \leq M_{max}$, the input soliton cannot display persistent rotary motion. In fact it performs a circular motion only at an initial stage of the propagation, subsequently transforming into a stationary stable state. A typical example of this evolution scenario is shown in Fig. 4(c). Obviously, this scenario is drastically different from those described above, when we considered lattices with sole refractive-index or loss modulation. Finally, in Fig. 4(d) we show a typical example of unstable evolution, leading to decay of the soliton.

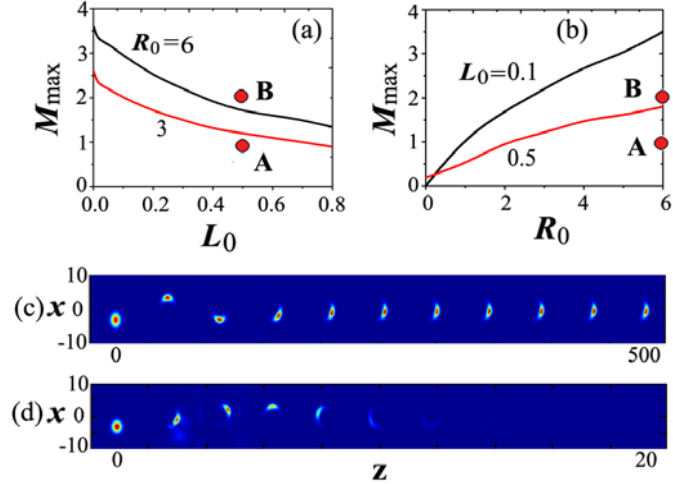

**Figure 4.** (Color online) The soliton's rotary motion in the first annular trough of the cylindrical lattices with the refractive-index modulation, $R(r) = R_0 \cos^2 r$, combined with the loss modulation, $L(r) = -L_0 \sin(2r)$. (a) The maximum strength of the tangential kick, $M_{max}$, versus loss modulation depth, $L_0$. (b) $M_{max}$ versus the refractive-index modulation depth, $R_0$. In (a) and (b), the motion is unstable above the solid curves; below the curves, the solitons perform stable rotary motion. (c) and (d): Stable and unstable evolution of the soliton at $M = 1$, $R_0 = 6$, $L_0 = 0.5$ and $M = 2$, $R_0 = 6$, $L_0 = 0.5$, respectively, corresponding to points A and B in (a) and (b). The input position of the soliton is at $x_{in} = \pi$. The results are presented as per Ref. [26].

## 6. Conclusions

In this article, we have reviewed recent results concerning the dynamics of dissipative spatial solitons in spatially modulated media described by the CQ CGL model. The following results have been highlighted.

First, we have studied 1D dissipative spatial solitons in the presence of sinusoidally modulated losses. It was found that the gradient force induced by the inhomogeneous loss gives rise to three generic propagation scenarios: (a) soliton's transverse drift; (b) persistent swing around the soliton's launching position; and (c) damped oscillations near or even far the input position.

Second, we have investigated the dynamics of 1D dissipative spatial solitons in the *PT*-symmetric periodic potentials. The unique properties found in this case are that the soliton can drift transversely or swing upon the propagation, under the action of the gradient force induced by the inhomogeneous loss. The drift or swing motion can be efficiently controlled by periods and amplitudes of the *PT*-symmetric periodic potential.

Third, we have studied the rotary motion of 2D dissipative spatial solitons in periodic cylindrical lattices of three types, which are generated by the refractive-index modulation, or linear-loss modulation, or a combination of both. Under the sole refractive-index modulation, or sole linear-loss modulation, the soliton exhibits stable rotary motion, provided that the initial tangential kick does not exceed a particular maximum value. For the combined refractive-index and linear-loss modulation, rotary motion of the soliton is transient, the soliton eventually halting at a certain spatial position. This effect does not occur in conservative counterparts of the system.

The obtained results suggest new possibilities for experimental and theoretical studies of the soliton dynamics, offered by complex lattice potentials. The results reported in this work are relevant because the generic CQ CGL model naturally arises in many physical settings, in addition to serving as the accurate model of the optical beam propagation in various lasing system – e.g., in semiconductor laser cavities with the loss modulation, and/or modulation of the refractive index.

A challenging direction for further studies is to consider similar possibilities in 3D models describing spatiotemporal localized modes in dissipative lattice media, i.e, dissipative "light bullets"; see, for example, Refs. [32-34].

**Acknowledgments**

This work was supported by the National Natural Science Foundation of China (Grant No. 11174061) and the Guangdong Province Natural Science Foundation of China (Grant No. S2013010015795). The work of D.M. was supported by CNCS-UEFISCDI, project number PN-II-ID-PCE-2011-3-0083.